\documentclass[preprint,preprintnumbers,amsmath,amssymb,superscriptaddress]{revtex4-1}
\usepackage{fullpage,amsthm,amsmath,amsfonts,bm,times,color,bbm}
\usepackage{graphicx,subfigure}
\usepackage[english]{babel}
\usepackage{textcomp}
\usepackage{mathtools}
\def\degree{\hbox{$^\circ$}}
\makeatletter

\begin{document}

\title{Climate network percolation reveals the expansion and weakening of the tropical component under global warming\\
}

\author{Jingfang Fan}
\email{j.fang.fan@gmail.com}
\affiliation{Department of Solar Energy and Environmental Physics, The Jacob Blaustein Institutes for Desert Research, Ben-Gurion University of the Negev, Midreshet Ben-Gurion 84990, Israel}
\affiliation{Department of Physics, Bar Ilan University, Ramat Gan 52900, Israel}

\author{Jun Meng}
\affiliation{Department of Solar Energy and Environmental Physics, The Jacob Blaustein Institutes for Desert Research, Ben-Gurion University of the Negev, Midreshet Ben-Gurion 84990, Israel}
\affiliation{Department of Physics, Bar Ilan University, Ramat Gan 52900, Israel}
\author{Yosef Ashkenazy}
\affiliation{Department of Solar Energy and Environmental Physics, The Jacob Blaustein Institutes for Desert Research, Ben-Gurion University of the Negev, Midreshet Ben-Gurion 84990, Israel}
\author{Shlomo Havlin}
\affiliation{Department of Physics, Bar Ilan University, Ramat Gan 52900, Israel}
\author{Hans Joachim Schellnhuber}
\affiliation{Potsdam Institute for Climate Impact Research, 14412 Potsdam, Germany}

\begin{abstract}
Global climate warming poses a significant challenge to humanity; it is associated with, e.g., rising sea level and declining Arctic sea ice. 
Increasing extreme events are also considered to be a result of climate warming~\cite{pachauri2014climate,ogorman_contrasting_2014} and they may have widespread and diverse effects on health, agriculture, economics and political conflicts \cite{hsiang2011civil,helbing_globally_2013,schleussner_armed-conflict_2016,carleton_social_2016}. Still, the detection and quantification of climate change, both in observations and climate models, constitute a main focus of the scientific community. Here we develop a new approach based on network and percolation frameworks \cite{cohen2010complex} to study the impacts of climate changes in the past decades using historical models and reanalysis records, and we analyze the expected upcoming impacts using various future global warming scenarios. 
More specifically, we classify the globe area into evolving percolation clusters and find a discontinuous phase transition, which indicates a consistent poleward expansion of the largest (tropical) cluster, as well as the weakening of the link's strength.
This is found both in the reanalysis data and in the Coupled Model Intercomparison Project Phase 5 (CMIP5) twenty-first century climate change simulations~\cite{taylor2012overview}. The analysis is based on high resolution surface (2~m) air temperature field records. We discuss the underlying mechanism for the observed expansion of the tropical cluster and associate it with changes in atmospheric circulation represented by the weakening and expansion of the Hadley cell.
Our framework can also be useful for forecasting  the extent of the tropical cluster in order to detect its influence on different areas 
in response to global warming.  
\end{abstract}

\maketitle
 
The series of powerful Atlantic hurricanes that hammered the Americas, the ``Lucifer'' heat wave that stifled Europe and the unusually dry June in Australia, all occurring in 2017, have been associated with the increased risks of extreme weather events \cite{Schellnhuber_2017}. Indeed, recent strong evidence supports the claim that this increase in extreme events is directly related to global warming \cite{Knutson2018}. Although some scientists question the significance of and the role played by human activity in global warming, the majority of the scientific community agrees that the warming is anthropogenic, due to elevated concentrations of heat-trapping (greenhouse) gases, especially triggered by the increased burning of fossil fuels and deforestation \cite{gleick2010climate}.

Network theory has demonstrated its potential as a useful tool for exploring the dynamical and structural properties of real-world systems from a wide variety of disciplines in physics, biology, and social science \cite{albert2002statistical,cohen2010complex,brockmann_hidden_2013,newman2010networks,romualdo_pastor-satorras_epidemic_2015,morone_influence_2015,gomez-gardenes_critical_2017}. Network approaches have been successfully implemented in climate sciences to construct ``climate networks,'' in which the geographical locations are regarded as network nodes, and the level of similarity between the climate records of different grid points represents the network links (strength). Climate networks have been successfully used  to analyze, model, and even predict climate phenomena \cite{yamasaki_climate_2008, ludescher_improved_2013, boers_prediction_2014, fan2017network}. Percolation theory has found to be an effective tool for understanding the resilience of connected clusters to node breakdowns through topological and structural properties \cite{bunde_fractals_1996,cohen_resilience_2000,aharony2003introduction}. The essence of the analysis is the identification of a system's different components and the connectivity between them. Percolation theory was applied to many natural and human-made systems \cite{taubert_global_2018, bunde_fractals_1996,li_percolation_2015,morone_influence_2015,meng_percolation_2017}. Here we combine climate network and percolation theory approaches to develop a framework with which to study and quantify the dynamical structure of the global climate system. Our results suggest that an abrupt first order percolation (phase) transition occurs during the evolution of the spatio-temporal climate networks. This evolution indicates the weakening and expansion of the giant component (tropical cluster) and is consistent with reported changes (expansion and weakening) of the tropical (Hadley) circulation.

\section*{Spatio-temporal climate networks}

Similar to earlier studies \cite{ludescher_improved_2013, fan2017network}, we construct a climate network based on the near-surface (monthly mean), high resolution ($0.125$\degree), air temperature of the ERA-Interim reanalysis data \cite{dee2011era} [Details are discussed in DATA AND METHODS]. We focus on the surface temperature field since it probably is the most commonly discussed global warming field; other variables, as well as other vertical layers, can be analyzed similarly to the way described below. Our evolving clustering process starts globally with $N = 1439\times2880$ isolate nodes (the South and North Pole grid points are eliminated). We then embed the network into a two-dimensional lattice where only nearest neighbor links are considered. The links are sorted in decreasing order of strength and then added one by one according to decreasing strength $W$ [Eq.~(\ref{eq4}) below]; i.e., we first choose the link with the highest weight, then the second strongest link and so on. More specifically, the nodes that are more similar (based on their temperature variations) are connected first. Existing clusters grow when a new link connects one cluster to another cluster (as small as a single node). We find that the climate network undergoes an abrupt and statistically significant phase transition, i.e., exhibiting a significant discontinuity in the order parameter $G_1$, the relative size of the largest cluster. Our results indicate that links with higher similarities tend to localize into a few large components (clusters of nodes), in the tropics and in the higher latitude regions (poles) of the Northern and Southern hemispheres. [We show the dynamical evolution of the climate networks in the supplementary animation~\cite{SI}.]

Fig.~\ref{Fig:1}(a) shows the climate network component (cluster) structure in
the globe map at the percolation threshold (just before the largest jump that is indicated by the orange arrow in Fig.~\ref{Fig:1}(b)). We find that the network, just before this jump, is characterized by three major communities; the largest one is located in the tropical region (indicated by red color); the second and third largest are located in the high latitudes of the southern (indicated by blue color) and northern (indicated by green color) hemispheres. In the next step, a critical bond will connect and merge the tropical cluster with the Southern Hemisphere cluster, resulting in a giant component. Fig.~\ref{Fig:1}(b) depicts the relative size of the largest cluster (the order parameter), $G_1$, as a function of the bond/link occupied probability $r$ in the evolution of the climate network. We find that $G_1$ exhibits an abrupt jump at the percolation threshold $r_c\approx 0.53$. The probability density function (PDF) of the weight $W_{i,j}$ of links is shown in Fig.~\ref{Fig:1}(d).

To study the significance of these results, we constructed the PDF of (temporally or spatially) reshuffled temperature records. In this way, either the memory within each record or the cross-correlations between the records are destroyed. We repeated the shuffling procedure 100 times and calculated their corresponding PDFs. The results, shown in Fig.~\ref{Fig:1}(b) and (d), suggest that in contrast to the randomly shuffled case, where the transition is continuous, the giant component in our real climate network shows an abrupt change.
The vertical line in Fig.~\ref{Fig:1}(d) indicates the strength of the critical link $W_c$ at the percolation threshold $r_c$. In addition, we find that $W_c$ is also  larger than the $95\%$ confidence level of the randomly shuffled data. 

During the growth of the climate network, we find, in Fig.~\ref{Fig:1}(b), that there are other smaller jumps. For example, the second largest jump occurs at $r\approx 0.8$, which is caused by the merging of the global cluster with the Northern Hemisphere high latitude cluster shown in blue in Fig.~\ref{Fig:1}(c).

\begin{figure}[!htb]
\includegraphics[width=1.0\linewidth]{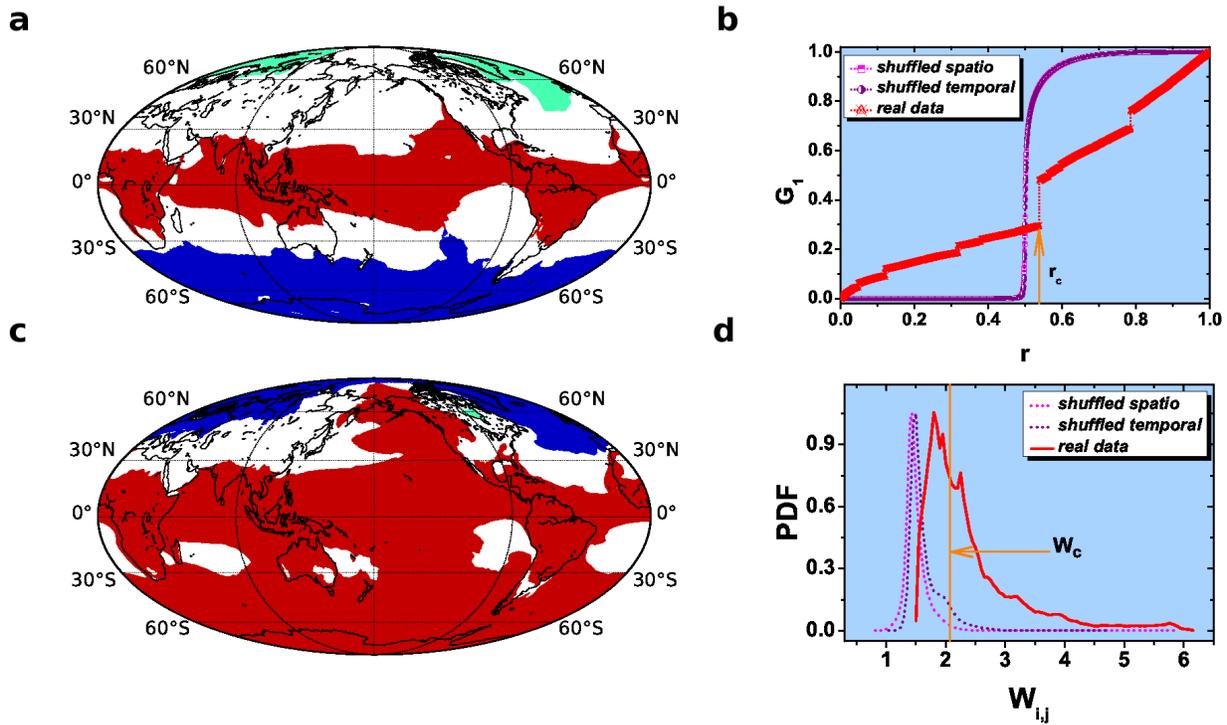}
\caption{\label{Fig:1}  Snapshots of the  component structures of the climate network. (a) Just before the percolation threshold (the largest jump  at $r\approx0.53$). Different colors represent different clusters; the grid resolution is 0.125\degree. (b) The giant cluster relative size $G_1$ versus the fraction number of bonds/links, $r$, for real (red), spatially shuffled (blue) and temporally shuffled (purple) records. (c) Just before the formation of the spanning cluster (at the second largest jump at $r\approx0.8$).   (d) The PDF of the weight of links $W_{i,j}$ around the globe, in real (solid line) and shuffled (dashed lines) data. The vertical line (orange) indicates the strength of the critical link, $W_c$, at the percolation threshold $r_c\approx0.53$; it is also higher than the $95\%$ confidence level of the randomly shuffled links.}
\end{figure}

\begin{figure}[!htb]
\begin{centering}
\includegraphics[width=1.0\linewidth]{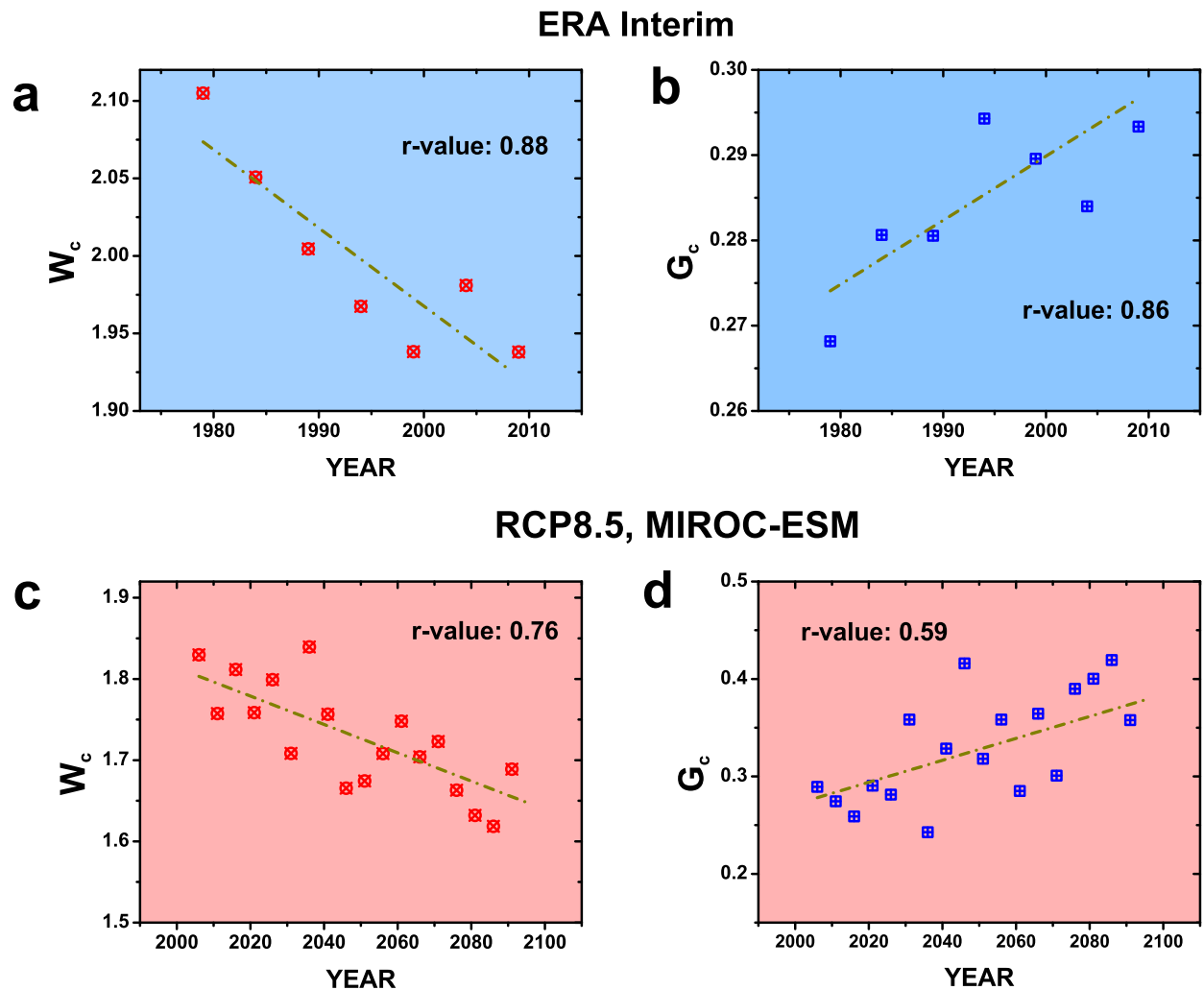}
\caption{\label{Fig:3} Weakening and expansion of the tropical component for (a-b) the ERA-Interim reanalysis data and for (c-d) a representative CMIP5 model, MIROC-ESM, under the RCP8.5 global warming scenario. $W_c$ is the weight of the critical link; $G_c$ is the normalized size of the tropical component just below  $r_c$, as shown in Fig. 1 (a) with red color.
Linear correlation coefficients (r-values) are given in the panels.}
\end{centering}
\end{figure}

To determine the temporal evolution of the size of the largest component $G_c$ (just before the largest jump at $r_c$), and its intensity $W_c$ (the weight of the critical link that leads to the largest transition; for more details on $G_c$ and $W_c$, see \textbf{DATA AND METHODS}), we construct a sequence of networks based on successive and non-overlapping temporal windows with lengths of $60$ months [5 years] each.  Fig~\ref{Fig:3} (a) depicts $W_c$ as a function of time indicating a significant decrease with time; $G_c$ shown in Fig~\ref{Fig:3} (b), however, exhibits an increasing trend with time. Specifically, by comparing the topological tropical component structures of the first and the last climate networks [shown in Fig. S1 in \cite{SI}], we find that the tropical cluster is expanding  poleward. This weakening and poleward expansion of the tropical component may be associated with global warming, as discussed below. An alternative definition of the cluster intensity is $\langle W \rangle$,
the average weight of links in the tropical cluster at the percolation threshold; this yields  similar results (see Fig. S2 in \cite{SI}). 

Next, we investigate the response of the tropical component to global warming using twenty-first century global warming experiments CMIP5 \cite{taylor2012overview}. We used the Representative Concentration Pathways 8.5 (RCP8.5) and 4.5 (RCP4.5), and Historical scenarios; the first two are future (twenty-first century) climate simulations under the assumption of warming by 8.5 and 4.5 watts, respectively, per meter squared. The results indicate a robust weakening and expansion of the tropical component for $26$ out of $31$ models in the RCP8.5 scenario [details are summarized in Table S1]. In Fig.~\ref{Fig:3}(c) and (d), we illustrate the changes of $W_c$ and $G_c$ with time, from 2006 to 2100 for one model, MIROC-ESM, under the RCP8.5 scenario. These $W_c$ and $G_c$ exhibit significant decreasing and increasing trends where the slope of the trends (i.e., the rate of change of $W_c$ and $G_c$ with time), $\xi_W$ and $\xi_G$, can be used to quantify the trend of each model [see Eq.~\ref{eq7}]. Fig.~\ref{Fig:4} (a), (b) and (c) show the results of  $\xi_W$ and $\xi_G$ for all 31 models, for the RCP8.5, RCP4.5 and Historical scenarios. We find that most of the models [$26/31$ (RCP8.5), $17/20$ (RCP4.5), $22/31$ (Historical)] show a stable trend and are located in the same phase [i.e., $\xi_W <0$ \& $\xi_G >0$]. The details (label number and resolution) of the 31 models we used are summarized in TABLE S1.

\section*{mechanism}
To study  a possible climatological  origin of the aforementioned results---weakening and expansion of the tropical cluster---we consider now the atmospheric circulation, especially, the Hadley cell (HC), which plays a pivotal role in the earth's climate by transporting energy and heat poleward. We conjecture that the tropical component of the climate network can be linked to the HC, since both of them exhibit consistent weakening and poleward expansion under global warming. In other words, the poleward expansion and weakening of the HC under global warming scenarios are linked to the expansion and weakening of the network-based tropical cluster. Below we present results that support this conjecture.

An analysis of satellite observations indicates a poleward expansion of the HC by $\sim 2$\degree~ latitude from 1979--2005 \cite{fu_enhanced_2006}. The main mechanisms for changes in the HC and its relation to global warming have yet to be elucidated \cite{held_nonlinear_1980}. A possible mechanism for the changes in the HC's and its relation to global warming has been proposed in ~\cite{lu_expansion_2007}. Also, a possible mechanism for the changes in the HC's strength and its relation to global warming was developed in  ~\cite{seo_mechanism_2014}. Both abovementioned observations and theories suggest a weakening and poleward expansion of the HC under global warming.

\begin{figure}[!htb]
\begin{centering}
\includegraphics[width=1.0\linewidth]{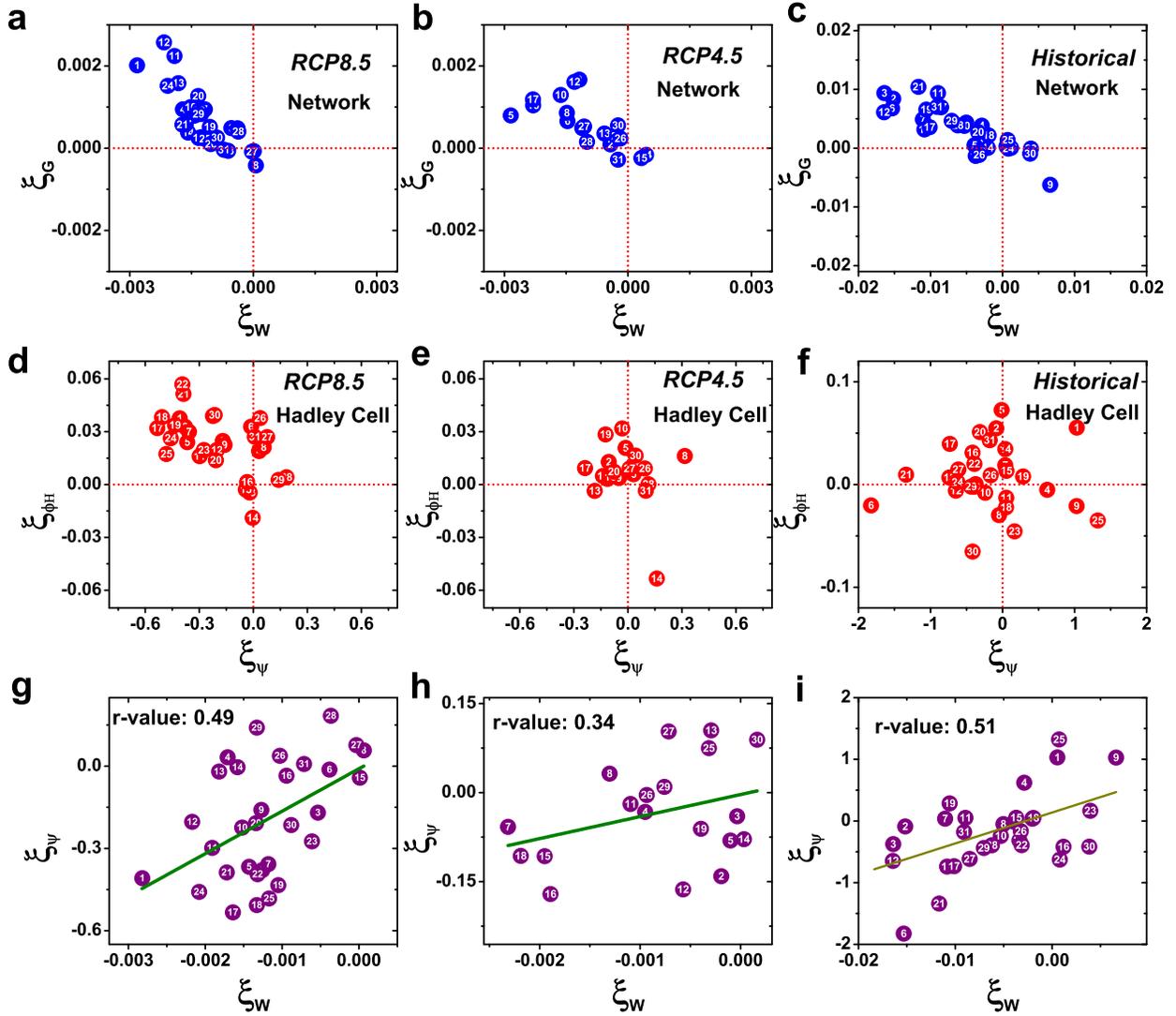}
\caption{\label{Fig:4} Changes in the size of the giant component against changes in the intensity for the network tropical component and the Hadley cell. The scatter plots of the increasing (size) and decreasing (intensity) trends for the (a, d) RCP8.5 scenario; (b, e) RCP4.5 scenario; and (c, f) Historical scenario. (g, h, i) Changes in the intensity of the HC against the network tropical component. Linear correlation coefficients ($r$ value) are given in the different panels. The numbers in the circles corresponding to the 31 models are summarized in TABLE S1.}
\end{centering}
\end{figure}

To find the relationship between the evolution of the climate network and the HC, we further calculate the stream function (Eq.~\ref{eq8}, see \cite{vallis2017atmospheric}) as a function of time, and from this, we evaluate the changes in the meridional width $\phi_H$ and strength $\Psi$ of the HC. We show the results in Fig. S3 for the ERA Interim reanalysis data and for a representative CMIP5 model under the RCP8.5 scenario. Similarly, we define $\xi_{\phi_H}$ and $\xi_\Psi$ as the change rates (the slope of the trend line) for the width and intensity of the HC. Fig.~\ref{Fig:4}(d), (e) and (f) show the corresponding results, where, as previously reported \cite{held_robust_2006,lu_expansion_2007,seo_mechanism_2014}, most of the CMIP5 models exhibit weakening and expansion of the HC [i.e., $\xi_\Psi <0$, $\xi_{\phi_H}>0$]. Fig. S4 depicts the results of a few CMIP5 models and indicates that there is a significant positive correlation between the tropical cluster intensity, $W_c$, and the intensity of the HC, $\Psi$. Similar correlations have also been observed between the width of the tropical cluster, $G_c$, and the width of the HC, $\phi_H$, for each individual model. A similar significant positive relationship has been found across all models by comparing $\xi_W$ and $\xi_\Psi$, see Fig.~\ref{Fig:4}(g), (h) and (i). Fig. S5 shows the positive relationship between the width of the tropical cluster and the HC. These results indicate that the tropical component of the climate network is correlated with the HC.

It should be noted that in contrast to model and theory ~\cite{held_robust_2006,seo_mechanism_2014}, an \textit{increasing} trend is found in the strength of HC in ERA Interim [Fig. S3(A)].
This is also confirmed by other NCEP--NCAR and ERA--40 reanalysis datasets \cite{mitas_has_2005}. A tentative hypothesis trying to resolve this contradiction was recently suggested \cite{held_robust_2006}. It is assumed that the increasing trends are artifacts related to the fact that the tropical lapse rate in the radiosonde data is increasing rather than staying close to a moist adiabat. However, researchers still have not found a way to deal with these artifacts. Our network-percolation approach seems to be more robust as it yields consistent \textit{decreasing} trend strength for the different datasets, including models and reanalysis, as well as different time periods.  This is  most likely due to the fact that the network-percolation approach is based on surface temperature data only. Fig. S6 shows our network results tested on the ERA--40 reanalysis---we find the same tendency of a weakening and an expansion of the tropical component under global warming.

\section*{Potential Effects}

\begin{figure}
\begin{centering}
\includegraphics[width=1.0\linewidth]{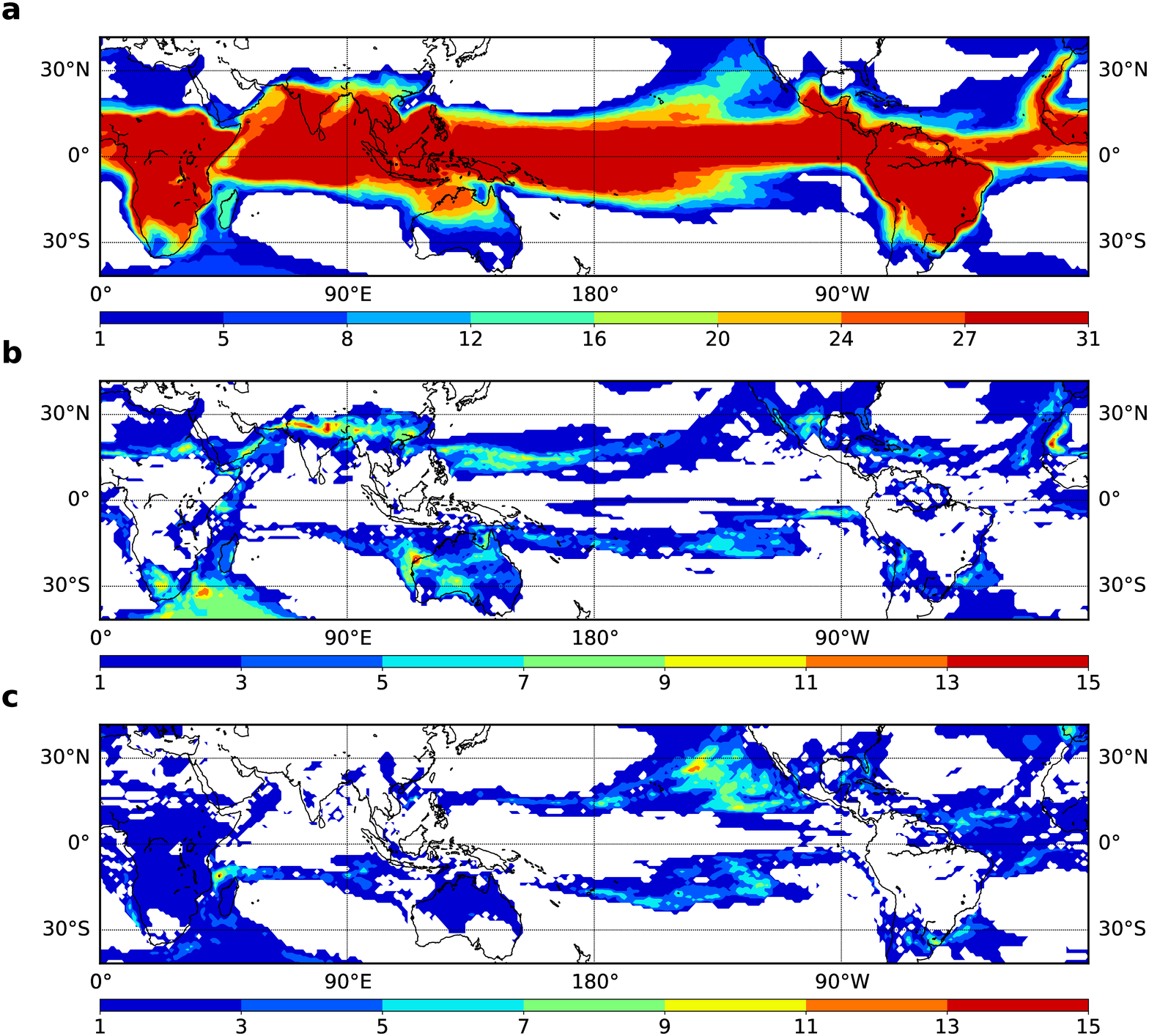}
\caption{\label{Fig:6} The evolution of the largest (tropical) cluster under the RCP8.5 scenario for the 31 models.  For each grid point (node), we compare the first and the last 20 years of the twenty-first century and show the number of models (a) with stable nodes (i.e.,  nodes that were part of the topical cluster for both the first and the last 20 years of the twenty-first century), (b) added nodes (i.e.,  nodes that were part of the tropical cluster for the last 20 years of the twenty-first century but not for the first 20 years), and (c) removed nodes (i.e.,  nodes that were part of the tropical cluster for the first 20 years of the twenty-first century but not for the last 20 years). }
\end{centering}
\end{figure}

The Intergovernmental Panel on Climate Change (IPCC) forecasts a temperature rise of 1.4 to 5.6 \textcelsius ~over this century (by 2100) \cite{pachauri2014climate}. Here, we associate the HC circulation with the tropical component found by the network-percolation approach. The locations of the subtropical dry zones and the major tropical/subtropical deserts are associated with the subsiding branches of the HC ~\cite{held_robust_2006,kelley_relative_2012}. Therefore, the poleward expansion of the HC may result in a drier future in some tropical/subtropical regions~\cite{lu_expansion_2007}. Another potential effect is the poleward migration of the location of tropical cyclone maximum intensity ~\cite{kossin_poleward_2014}. There is thus a great interest in clarifying/predicting the influenced   climatological areas in response to global warming.

To identify the climate change response, for simplicity but without loss of generality, we compare, in Fig. \ref{Fig:6}, the topology of the tropical component for the first and last twenty years of the twenty-first century, i.e., (2080--2100) vs. (2006--2026). Counting the number of models that simulate these changes (Fig.~\ref{Fig:6}), one can see that the overall pattern of stable node change is robust across most models; the patterns of adding nodes or removing nodes are shown in Fig.~\ref{Fig:6}(b) and (c). We find that some regions, for example, northern India, southern Africa and western  Australia have a higher probability to be influenced by the tropical component (or HC), whereas the impacts in other regions, e.g., the Northeast Pacific, will become weaker in the future. For ease of comparison, the datasets are interpolated into a $1$\degree$\times$ $1$\degree longitude-latitude grid.

\section*{Discussion}
It has been pointed out that a random network or lattice system always undergoes a continuous percolation phase transition and shows standard scaling features during a random process ~\cite{bunde_fractals_1996,bollobas2001random}. The question whether percolation transitions could be discontinuous has attracted much attention recently in the context of interdependent networks~\cite{buldyrev_catastrophic_2010} and the so-called explosive percolation models ~\cite{achlioptas_explosive_2009, riordan_explosive_2011, dsouza_anomalous_2015}. Interestingly, the dynamic evolution of our climate network indicates the possibility of discontinuous phase transition, as shown in Fig~\ref{Fig:1}(b). To further test the order of the percolation phase transitions, we study the finite size effects of our network by altering the resolution of nodes. The results are shown in Fig. S7. It suggests a discontinuous percolation. For comparison, we also show the results for shuffled data (shuffled spatial), which exhibit a continuous phase transition.

In summary, we used a network and percolation analysis on surface air temperature data and found that the largest (tropical) cluster expands poleward and experiences weakening under the influence of global warming. We show that these trends are significant, and we relate them to the weakening and poleward expansion of the HC atmospheric circulation. By comparing the topology of the tropical component for the first and last 20 years of the twenty-first century, i.e., (2080--2100) vs. (2006--2026), we clarify/predict the influenced climatological areas in response to global warming. Furthermore, we find an abrupt jump during the dynamical evolution of the climate network; using a finite size scaling analysis, we argue that the percolation transition is first order. The study of the climate system may enrich the understanding of the discontinuous phase transition. The proposed method and analysis provide a new perspective on global warming and can potentially be used as a template to study other climate change phenomena.

\section*{Data and Methods}
\subsection*{Data}
In this study, we employ the monthly 2 m near surface air temperature and 37 pressure level meridional wind velocity $V$ of ERA--Interim \cite{dee2011era} and ERA--40 \cite{uppala_era-40_2005} reanalysis datasets. The resolution is $0.125$\degree, the time period spans from 1979 to 2016 for ERA--Interim and 1980 to 2005 for ERA--40. The data can be downloaded from \url{http://apps.ecmwf.int/datasets/}.

For the climate projection, we used the RCP8.5, RCP4.5 and Historical scenarios of the CMIP5. RCP8.5 is the upper bound of the RCPs and does not include any specific climate mitigation target. The RCP8.5 assumes that greenhouse gas emissions and concentrations increase considerably over time, leading to a radiative forcing that will stabilize at about 8.5 $W m^{-2}$ at the end of the twenty-first century. All analyzed data are monthly mean surface air (2~m) temperatures and the meridional component of wind, $V$. The details of all 31 models we used are summarized in Table S1~\cite{SI}. The data can be downloaded in \url{http://pcmdi9.llnl.gov/}.

\subsection*{Complex networks}
For each node $i$ (i.e., longitude-latitude grid point), given a temperature record $\tilde{T}_{i}(t)$, where $t$ stands for the month, we consider the month-to-month temperature difference 
\begin{equation}
T_{i}(t) =  \tilde{T}_{i}(t+1) - \tilde{T}_{i}(t).
\label{eq1}
\end{equation}
We take the difference since we focus on climate change and how a change in one node affects other nodes; we repeated the analysis using the original time series and obtained similar clusters. 

To obtain the strengths of the links between each pair of nodes $i$ and $j$, we define the time-delayed cross-correlation function as,
\begin{equation}
C_{i,j}(\tau) =  \frac{<T_{i}(t)T_{j}(t+\tau)> -<T_{i}(t)><T_{j}(t+\tau)>}{\sqrt{(T_{i}(t) - <T_{i}(t)>)^2} \cdot \sqrt{(<T_{j}(t+\tau) - <T_{j}(t+\tau)>>)^2}},
\label{eq2}
\end{equation}
and
\begin{equation}
C_{i,j}(-\tau) = C_{j,i}(\tau),
\label{eq3}
\end{equation}
where  $\tau$ is the time lag between 0 and 24 months; for the networks that span a short time of $5$ years, we set $\tau\in[0, 12]$. We define the strength of the link as
\begin{equation}
W_{i,j} = \frac{\max(|C_{i,j}( \tau)|) - {\rm mean}(C_{i,j}(\tau))}{{\rm std}(C_{i,j}(\tau))},
\label{eq4}
\end{equation}
where ``max'', ``mean'' and ``std'' are the maximum, mean and standard deviations of the cross-correlation ~\cite{fan2017network}.

Based on classical graph theory, a component is a subset of network nodes such that there exists at least one path from each node in the subset to another~\cite{cohen2010complex,newman2010networks}. We denote $S_m$ as a series of sub-networks; specifically, $S_1$ indicates the largest cluster, $S_2$ indicates the second largest cluster, and so forth. In this study, due to the earth’s spherical shape, the largest component in the climate networks is defined as
\begin{equation}
G_1 = \frac{\max \left[\sum\limits_{i\in S_1} \cos(\phi_i),\cdots, \sum\limits_{i\in S_m} \cos(\phi_i),\cdots,\right]}{\sum\limits_{i=1}^{N} \cos(\phi_i)}, 
\label{eq5}
\end{equation}
where $\phi_i$ is the latitude of node $i$. Since our network is finite, we use the following procedure to determine the percolation threshold. We first calculate, during the growth process, the largest change of $G_1$,
\begin{equation}
\Delta \equiv \max\left[G_1(2)-G_1(1),\cdots,G_1(M+1)-G_1(M),\cdots\right].
\label{eq6}
\end{equation}
where $M$ is the number of links  (ordered by decreasing weight value) we added. 
The step with the largest jump is regarded as the phase transition point. We consider and analyze the weight of the critical link $W_c$ and the size of the largest component $G_c$, which represent the intensity and size of the component. To measure the change of $W_c$ and $G_c$ within time, we anticipate the slope of the trend lines 
\begin{align}
W_c (t) & = a + \xi_W *t  \nonumber \\
G_c (t) & = b + \xi_G *t,  
\label{eq7}
\end{align}
where $\xi_W$ and $\xi_G$ denote the increasing or decreasing trend rate, $a$ and $b$ are constants, and $t$ is the time; we use a time interval of five years.

\subsection*{Hadley cell index}
The strength of the HC is computed using observed zonal-mean meridional wind in the stream function $\Psi$~ \cite{vallis2017atmospheric},
\begin{equation}
[\overline{V}] =  \frac{g}{2\pi R \cos\phi} \frac{\partial \Psi}{\partial p}, 
\label{eq8}
\end{equation}
where $V$ is the meridional velocity in pressure coordinates, $R$ is the mean radius of the earth, and $p$ is the pressure. The operators $\bar{ }$ and $[ ~]$ stand for temporal and zonal averaging, respectively. We compute the $\Psi$ field, assuming $\Psi = 0$ at the top of the atmosphere and integrating Eq.~\ref{eq8} downward to the surface. Since in the winter, the HC is stronger, we only focus on the winter HC intensity. The analyses are performed on December-January-February (DJF) for the Northern Hemisphere (NH) and June-July-August (JJA) for the Southern Hemisphere (SH) separately. Fig. S8 shows the mean stream function $\Psi$ based on ERA--Interim during DJF and JJA.

Then we denote the maximum (minimum, respectively) of $\Psi$ as $\Psi_{N}$ ($\Psi_{S}$) during DJF (JJA)  over the tropics [-20\degree, 20\degree];  the corresponding pressure level is denoted as $p_N$ ($p_S$). The HC strength is defined as the difference between the values of the maximum and minimum, $\Psi = \Psi_{N} - \Psi_{S}$. 

We identified the northern (southern) boundary $\phi_N$ ($\phi_S$) of the HC as the first latitude north (south) poleward of $\Psi_{N}$ ($\Psi_{S}$) at $p_N$ ($p_S$) becomes zero. The poleward edges of the HC are defined as $\phi_H = \phi_N - \phi_S$.  To measure the change of $\Psi$ and $\phi_H$ within time, we define $\xi_\Psi$ and $\xi_{\phi_{H}}$, analogous to Eq.~\ref{eq7}.

\section*{Acknowledgements}
We thank Ori Adam and Maaian Rotstein for helping us to retrieve the CMIP5 data and Jianxi Gao, Daqing Li and Xiaosong Chen for helpful discussions. We acknowledge the Israel-Italian collaborative project NECST, the Israel Science Foundation, ONR, the Japan Science Foundation, BSF-NSF, and DTRA (Grant No. HDTRA-1-10-1-0014) for financial support.

\section*{Contributions} All authors designed the research, analyzed data, discussed results, and contributed to writing the manuscript.

\section*{Additional information}
Supplementary information is available in the online version of the paper. 

\section*{Competing financial interests}
The authors declare no competing financial interests

\bibliographystyle{naturemag}

\bibliography{MyLibrary}

\end{document}